\documentclass[twocolumn,aps,prb,showpacs]{revtex4}

\usepackage{graphicx}
\usepackage[usenames]{color}
\usepackage{amsmath}
\usepackage{amssymb}
\usepackage{subfigure}
\usepackage{placeins}
\usepackage{amssymb}

\def\up{\uparrow}
\def\dn{\downarrow}

\def\ot{\frac{1}{3}}
\def\tt{\frac{2}{3}}
\def\tf{\frac{2}{5}}

\def\vek#1{\vec{#1}}

\def\ve{\varepsilon}
\def\enu{e^2/(4\pi\ve\ell_0)}

\begin{document}
\title{Response of incompressible fractional quantum Hall states to
  magnetic and non-magnetic impurities}

\author{Karel V\'yborn\'y}
\affiliation{Fyzik\'aln\'\i{} \'ustav, Akademie v\v ed \v Cesk\'e republiky,
  Cukrovarnick\'a 10, Praha 6, CZ--16253, Czech Republic}

\author{Christian M\"uller}
\author{Daniela Pfannkuche}
\affiliation{1. Institut f\"ur theoretische Physik, Universit\"at
  Hamburg, Jungiusstr. 9, D--22305 Hamburg, Germany}

\date{March 4th, 2007}

\begin{abstract}
Using exact diagonalization we examine the response of several most
prominent fractional quantum Hall states to a single local
impurity. The $2/3$ singlet state is found to be more inert than the
polarized one in spite of its smaller incompressibility gap. 
Based on its spin-spin correlation functions we interpret it as a
liquid of electron pairs with opposite spin. A comparison of different
types of impurities, non--magnetic and magnetic, is presented.
\end{abstract}

\maketitle

The effect of impurities on the incompressible ground states was in the
focus of many theoretical studies since the early days of the
fractional quantum Hall effect (FQHE). The principal interest is based
on the knowledge that the FQHE occurs only in high mobility samples
implying that random electric potential ubiquitous in experiments
leads to a decrease of the incompressibility gap eventually
destroying the incompressibility completely. A large part of the
theoretical work therefore deals with random ensembles of
impurities\cite{haldane2005}
and this, due to the complexity of such calculations, in context of
the most robust incompressible FQH state, bearing the name of
R.~Laughlin, which occurs at filling factor\cite{chakraborty:1995}
$\nu=\ot$.

A single impurity can, however, be also viewed as a probe into the correlated
state revealing additional information on its properties and
nature\cite{martin:08:2004}. Beyond this, as we believe fundamental interest,
this information may help to understand new correlated states for which only
numerical many-body wavefunctions are available. The particular example we
bear in mind are the half-polarized states at $\nu=\tt$ and $\tf$ discovered
first in optical experiments\cite{kukushkin:05:1999}. Numerically, the
wavefunctions of suitable candidates can be obtained by exact diagonalization
but it was not possible to link them conclusively with any of the physical
states proposed
subsequently\cite{vyborny:01:2007,mariani:12:2002,merlo:04:2005}%
\cite{apalkov:02:2001,murthy:01:2000}.
The current study of the parent polarized and unpolarized states may provide a
guideline to achieve this.

First, the Laughlin wavefunction is investigated and different regimes
regarding tis response to a point impurity are identified.  Here, we build on
the seminal articles of Zhang\cite{zhang:11:1985} and
Rezayi\cite{rezayi:11:1985}. A comparison between point and $\delta$-line
impurities, is presented.  Proceeding to filling factors $\tt$ and $\tf$, it
proves advantageous to apply different types of impurities, magnetic and
non-magnetic.  This investigation reveals basic structural properties of the
various FQH states like local density and polarization responses to
impurities. Based on these, we propose that the $\nu=\tt$ singlet state is a
$\nu=1$ liquid of electron pairs with opposite spin.

We recall that the three filling factors studied here, $\ot$, $\tt$
and $\tf$ are all closely related. In the composite fermion (CF)
picture\cite{heinonenbook}, they correspond to filling factors $\nu^*$
equal to 1, 2 and 2 respectively and the last two differ only in the
direction of the effective field.  Systems at $\nu=\tt$ and $\tf$ can
therefore be expected to behave very similarly, at least within the mean
field of the CF. On the other hand, with the spin
degree of freedom frozen out (e.g. by large Zeeman energy), the ground
states at $\nu=\ot$ and $\tt$ are identical up to a particle-hole
conjugation\cite{dev:11:1992,vyborny:2005}. Here we show how far these
statements of close relationship apply when the excitations become
involved in addition to the ground states as it is the case with
inhomogeneous systems.

\section{The model and point impurities}

\begin{figure}
\begin{tabular}{cc}
\hskip-.5cm \includegraphics[scale=0.45]{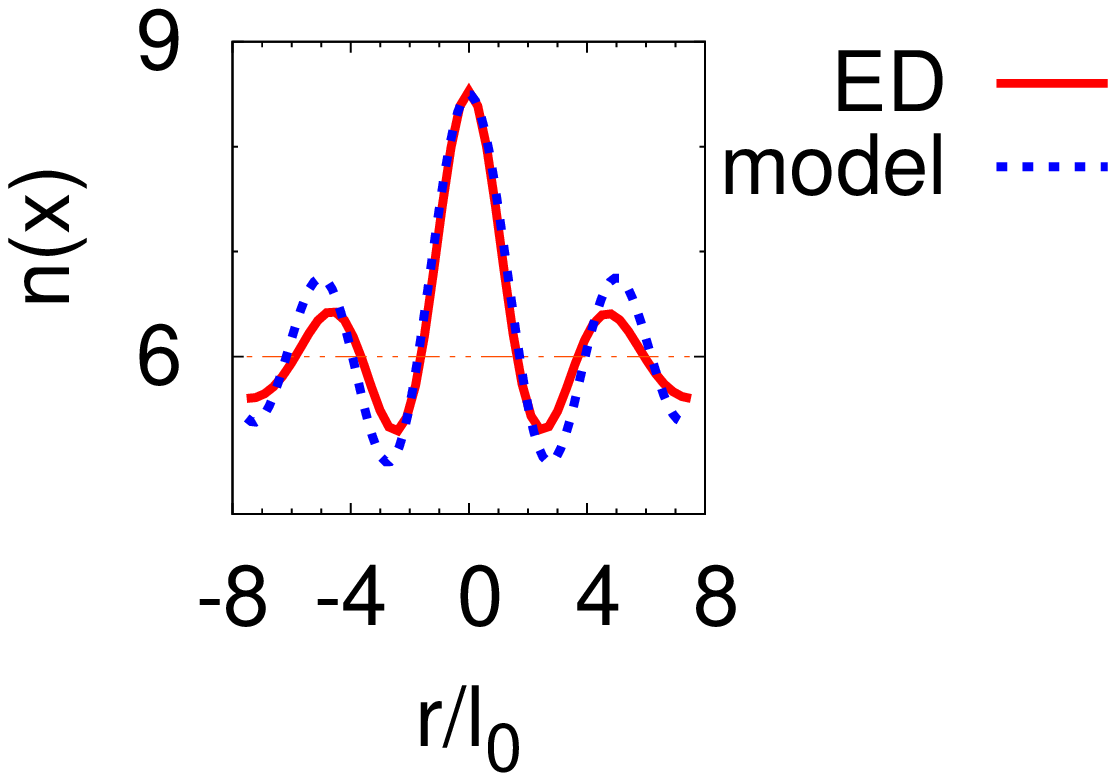} &
\hskip-2cm \raise-5mm\hbox{\includegraphics[scale=0.55]{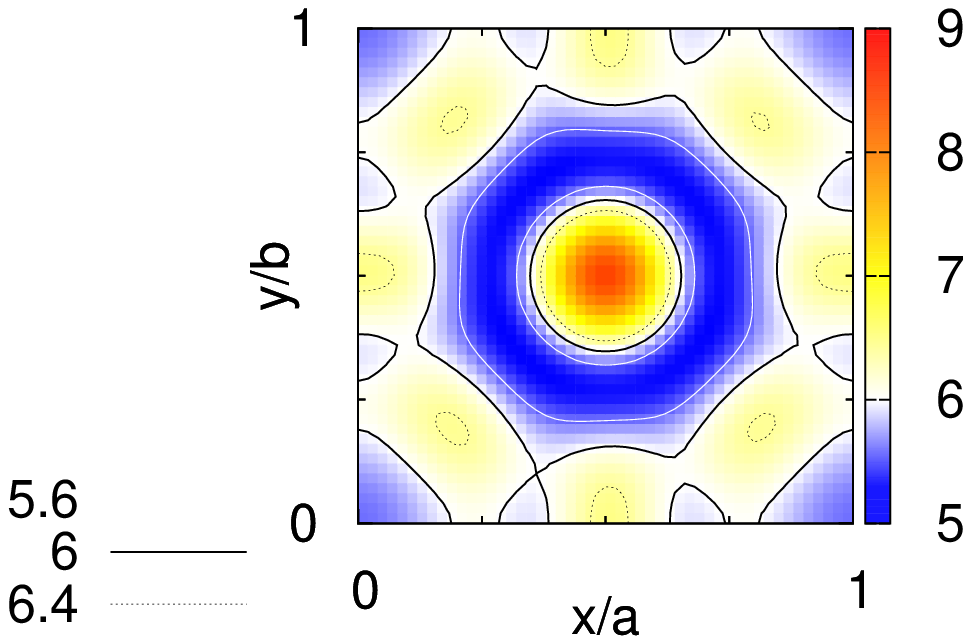}} \\
(a) & (b)
\end{tabular}
\caption{{\em Right:} electron density $n(x,y)$ of the Laughlin state
  in a square with periodic boundary conditions under the influence of
  a single point impurity of intermediate strength. Six electrons
  (implying $a=b=10.6\ell_0$ to ensure $\nu=\ot$), $V_0=0.3\enu$,
  $\sigma=1.0\ell_0$ (Eq.~\ref{eq-02}). {\em Left:} section of
  $n(x,y)$ along the diagonal (ED), comparison to the simple model
  based on $\chi(\vec{q})=\delta(|\vec{q}-q_0|)$, $q_0\ell_0=1.4$, see
  text.}  
\label{fig-14}
\end{figure}

It is a common notion that the Laughlin state is incompressible. This
statement however relates to a thermodynamical property and does not
contradict the fact that even an arbitrarily small impurity potential
locally changes the electron density, Fig.~\ref{fig-14}. 

\begin{figure}
\includegraphics[scale=.6]{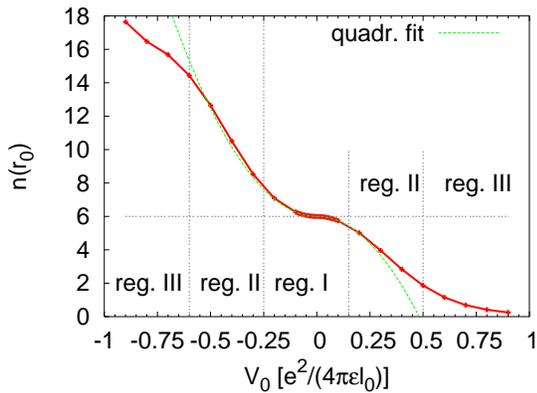}
\caption{Compression of the Laughlin state by a point impurity
  measured by the local density at $r_0$ (Eq.~\ref{eq-02}). Strength
  of the impurity $V_0$ is divided into regions I, II and III
  described in text.}
\label{fig-21}
\end{figure}

As a function of the impurity strength $V_0$, the local density at the
position of the impurity $n(r_0)$ shows three distinct types of
behaviour marked by I, II and III in Fig.~\ref{fig-21}. The boundary
between I and II can roughly be identified with the incompressibility
gap $E_g$. While the density change $\Delta n(r_0)$ is roughly
proportional to $V_0$ for $|V_0|\gtrsim E_g$ (region II), reminiscent of a
standard compressible behaviour as of a Fermi gas or liquid, the
density change is proportional to $V_0^2$ for weak impurities (region I). 
The latter non-linear region does not appear in earlier
data\cite{rezayi:11:1985} on a sphere and we attribute it to the
center-of-mass part of the wavefunction as we explain below. 
Finally, the response diverts again from the linear
regime for very strong impurities (region III). We will not
investigate this regime here and focus only on the regions of $\Delta
n(r_0)\propto V_0$ and $|\Delta n(r_0)|\propto V_0^2$.

As a tool of study we use the exact diagonalization (ED) with 
Coulomb-interacting electrons on
a torus\cite{chakraborty:1995,yoshioka:2002,yoshioka:06:1984}. Contrary
to the spherical geometry, a point impurity
\begin{equation}\label{eq-02}
V({\vec{r}})= V_{0}\exp{[-({\vec{r}}-{\vec{r}_{0}})^2/\sigma^2]}
\end{equation}
does break geometric symmetries on the torus described by the quantum
numbers\cite{haldane:11:1985} $\mathbf{k}=(k_x,k_y)$. This makes the study
more difficult from the computational view (large dimension of the
Hilbert space) but it also gives the system the full freedom of choosing
the ground state. The density response $n(x,y)$, Fig.~\ref{fig-14}, clearly
follows the impurity form (rotationally symmetric) on short ranges and
it is deformed by the periodic boundary conditions on distances
comparable to the size of the elementary cell\cite{chakraborty:1995}
$a=b=\sqrt{2\pi N_m}\ell_0$. Here $\ell_0$ is the magnetic length and
$N_e$, $N_m$ are the number of electrons and the number of magnetic
flux quanta. 

It should be noted that the oscillations in the density response
$n(x,y)$ are not of the Friedel type known in a Fermi
gas\cite{ashcroft:1976,stern:04:1967,stern:11:1967} which occur as
interferences at the edge of the Fermi--Dirac distribution at the
Fermi wave vector. In the regime of linear response, the oscillations
like in Fig.~\ref{fig-21} can be described up to a very good precision
by a model dielectric response function
$\chi(\vec{q})=\delta(|\vec{q}|-q_0)$. This is an approximation based
on the observation that the magnetoroton minimum at $q_0\ell_0\approx
1.4$ is the lowest no-spinflip excitation at $\nu=\ot$. A more
realistic $\chi(\vec{q})$ based on the single mode approximation was
given by MacDonald\cite{macdonald:03:1986} and we recall that it is
very different from $\chi(\vec{q})$ of a Fermi
gas\cite{ashcroft:1976,macdonald:03:1986}.

It was noted already by Rezayi and Haldane\cite{rezayi:11:1985} that the form
of the density response calculated for the linear regime, $\Delta n(r)\propto
J_0(rq_0)$, remains almost unchanged even in the non-linear regime $|V_0|\gg
E_g$. This statement applies also for point impurities on a torus with two
comments.  (a) The density profile predicted by the linear response
calculation is correct also outside the linear regime (large $V_0$), but not
all the way up to $V_0\to \infty$ as on a sphere\cite{rezayi:11:1985}. (b)
This density profile $n$ is also correct when $V_0$ drops into the $\Delta
n(r_0)\propto V_0^2$ regime. However, it may be masked by density modulation
due to finite size effects in small systems\cite{vyborny:2005}.



On a torus at at $\nu=1/3$, the CM
wavefunctions span a three-dimensional space\cite{haldane:11:1985}. States in
a homogeneous system can be factorized into a CM and relative parts,
$\Psi=\Psi_{CM}\psi_{rel}$. Even the incompressible Laughlin ground state is
thus triply degenerate on the torus and the electron density in this state
strongly depends on which linear combination we choose for its $\Psi_{CM}$.
Impurities lift the degeneracy but the splitting $\Delta E$ remains very
small, $\Delta E\ll V_0\lesssim E_g$. Density responses in each state $\Delta
n_{V_0}=n_{V_0}-n_{V_0\to 0}$ are not identical but differences and vanish
with increasing system size. The roles of the three states can be interchanged
by moving $\vec{r}_0$, Eq.~\ref{eq-02}. We always chose the state with lowest
energy for plots in this article.

We attribute the $\Delta n(r_0)\propto V_0^2$ behaviour to the situation where
the CM part of the wavefunction enters as a degree of freedom. Matrix elements
of the electron density between two states are proportional to the overlap of
their CM parts. In particular, if these are mutually orthogonal, the
corrections to the density linear in $V_0$ will vanish. For stronger impurity
potentials, the CM degree of freedom may be frozen out and linear admixtures
to the ground state wavefunction imply linear corrections to the electron
density. For electrons on a sphere or impurities discussed in the next section,
the CM degree of freedom is never enabled and the density response is linear
down to $V_0\to 0$.

\section{Response to a $\delta$-line impurity}

Comparing results for different system sizes is essential in finite
system studies. To keep such calculations tractable we must return to
impurities which preserve some symmetries of the system. The
$\delta$-line impurity
$$
  V_{\delta\mbox{--line}}(x,y)=V_0\delta(x-x_0)\,,
$$
in a rectangle with periodic boundary conditions has a similar
position as a point impurity on a sphere. The former conserves
translational symmetry along $y$, the latter conserves projection of
the angular momentum to one axis. A straightforward consequence is
that a $\delta$-line impurity does not mix the center-of-mass
degenerate ground states on a torus. From the viewpoint of perturbation
theory, the ground states are then non-degenerate and the regime of 
$\Delta n(r_0)\propto V_0^2$, Fig.~\ref{fig-21}, is thus absent.
Consequently, $\Delta n(r_0)$ is linear in $V_0$ even for $|V_0|\lesssim E_g$.

Throughout the rest of the article we will only use the $\delta$--line
impurity with a strength of $V_0=0.01\enu$ which is for all studied states
$\ll E_g$. Calculated responses then provide information on the relative
robustness of the different states (note however that the response is not a
property of the ground state alone but depends also on low excited states).
The form of the response is found to be independent on $V_0$ as long as
$|V_0|\lesssim E_g$. Potentially, $V_{\delta\mbox{--line}}$ may be useful in
studies of domain walls but we do not follow this aim in this
article\cite{vyborny:2005}.

\begin{figure}
\includegraphics[scale=.6]{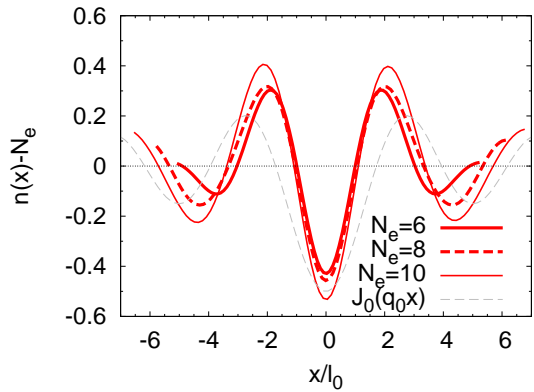}
\caption{The Laughlin state responding to a $\delta$-line repulsive
  impurity. Responses in systems of different sizes are compared among
  each other and also to the linear response
  model\cite{rezayi:11:1985} for {\em point} impurities,
  cf. Fig.~\ref{fig-14}.}
\label{fig-01}
\end{figure}

Turning to the $\ot$ Laughlin state, a linear-response analysis as in
Ref.~\cite{rezayi:11:1985} yields $\Delta n(x,y)\propto
\exp(-x^2/s^2)$ for $x_0=0$.  It is noteworthy that the form of
$\Delta n(x,y)$ should be now governed by the width $s$ of the peak of
$\chi(\vec{q})$ rather than by its position $q_0$.  However, this
simple model of $\chi(\vec{q})$ seems to fail as the Laughlin state
density response is again oscillatory, Fig.~\ref{fig-01}, yet with different
period than for point impurities (dotted line).

Keeping in mind the spin degree of freedom, an impurity can
principally be one of the following three different types
($\delta_{\sigma\up}$ is a projector to states with spin $\up$)
\begin{equation}\label{eq-01}
  H_{imp}=E_{imp}\sum_{i=1}^{N_e} V(\vek{r}_i) \otimes \left\{ 
    \begin{array}{ll}\delta_{\sigma_i\up}+\delta_{\sigma_i\dn} &
      \mbox{ 'EI'}\\
           \delta_{\sigma_i\up}-\delta_{\sigma_i\dn} &
      \mbox{ 'MI'}\\
           \delta_{\sigma_i\up} &
      \mbox{ 'DP'}\end{array}\right.\,.
\end{equation}
The electric potential impurity (EI) stands for a simple potential modulation
($\delta_{\sigma_i\up}+\delta_{\sigma_i\dn}=1$), the magnetic
impurity (MI) can be viewed as a spatially varying Zeeman field
[$V(\vek{r}_i)\otimes(\delta_{\sigma_i\up}-\delta_{\sigma_i\dn}) \propto
  B_z(\vek{r}_i)\sigma^z_i$] and the delta--plus impurity (DP) acts as a
potential modulation seen only by spin--up electrons. The last type of
impurity, DP, is less likely to occur in physical systems, however, it
is helpful to understand the mechanisms governing inner structure of
the states in study.

It is important to check whether the response does not decay with
increasing system size. Such a behaviour may imply that the response
vanishes in large enough systems. However, in most of cases we find even a
slight increase of the density response $\Delta n(r)$,
e.g. Fig.~\ref{fig-21}. Exceptions from this will be explicitly stated.

\subsection{Electric potential impurity (EI)}

The polarized states of $\ot,\tt$ and $\tf$ respond all very similarly
to a weak potential impurity. The singlet states at $\tf$ and $\tt$
respond more weakly and this is particularly apparent in the latter
case. We give details on this observation in this Section and then
focus on the $\tt$ singlet state in the succeeding Sections.

\begin{figure}
\includegraphics[scale=.6]{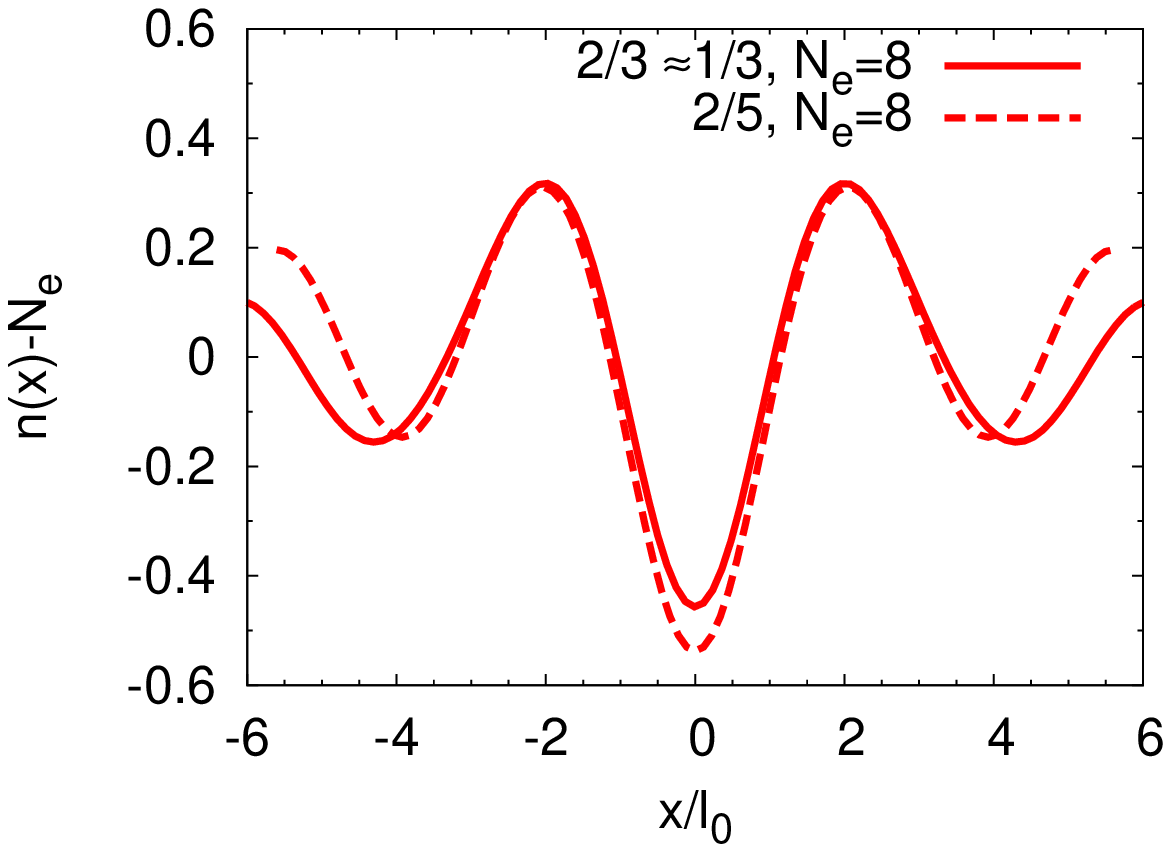}
\caption{Response of the polarized $\tf$ and $\tt$ (almost identical
  to that of $\ot$) states to a $\delta$-line EI. Eight
  electrons considered, $V_0=0.01\enu$.}
\label{fig-03}
\end{figure}

It is not surprising that the Laughlin state and the polarized state
at $\tt$ have a similar response. They are particle hole conjugates
($\nu$ and $1-\nu$) and even though inhomogeneities break the symmetry
of the Hamiltonian\cite{vyborny:2005}, the effect appears extremely
weak on the scale of Fig.~\ref{fig-03}. Also the $\tf$ polarized state and
the Laughlin state show almost the same response, Fig.~\ref{fig-03}. 
This applies both to its strength and the position of the first
maximum (or node). Such a finding is non-trivial as the two states
correspond to different filling factors in the CF picture. It is also
non-trivial from the analogy between $\tf$ and $\tt$. Here both states
correspond to $\nu^*=2$ but the wavefunctions show different
correlations on the electronic level\cite{vyborny:2005}.

\begin{figure}
\begin{tabular}{cc}
\hskip-0.3cm\includegraphics[scale=.45]{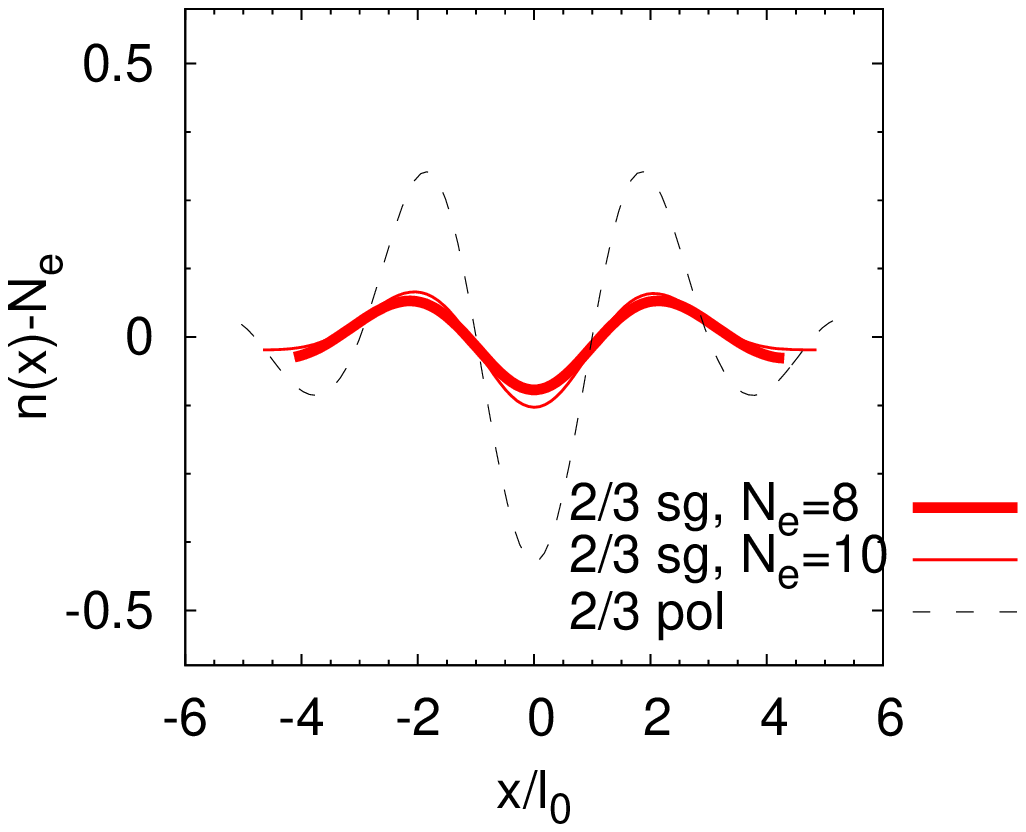} &
\hskip-1.3cm\includegraphics[scale=.45]{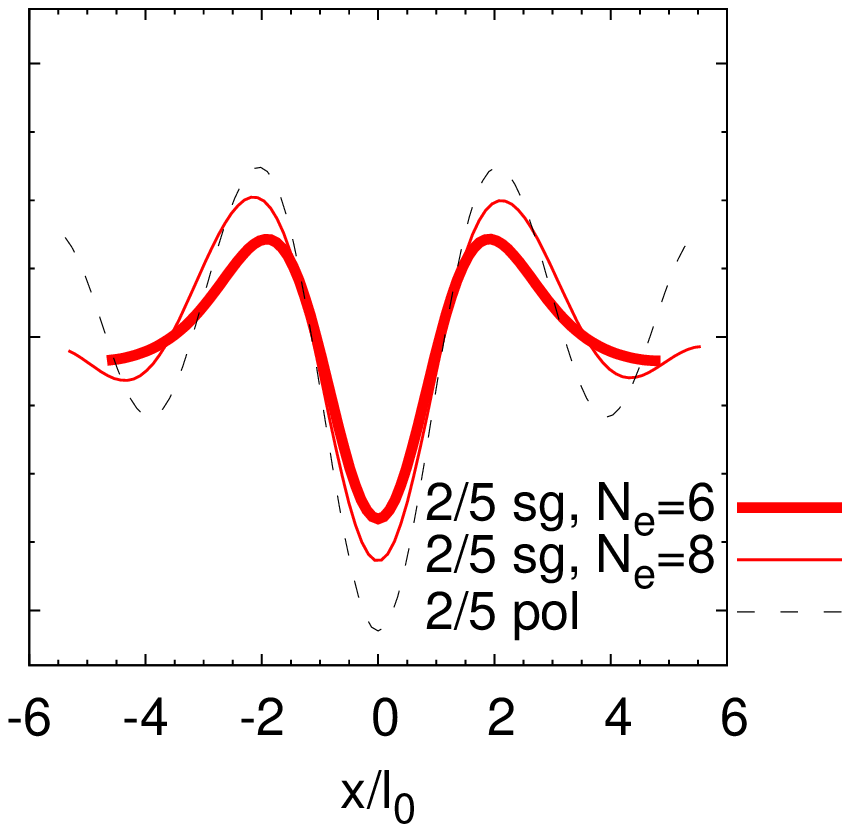}\\
(a) $\tt$ singlet &
(b) $\tf$ singlet
\end{tabular}
\caption{The singlet states subject to a potential impurity. Response
  of the polarized states (thin dashed lines) is shown for comparison.}
\label{fig-04}
\end{figure}

Now turn to the {\em singlet} GS, Fig.~\ref{fig-04}. The $\tf$ singlet
seems to respond less strongly compared to the polarized state but the
difference is small, Fig.~\ref{fig-04}b (crosses and the thin line). 
The difference between the polarized and singlet state at
$\tt$, however, is striking, Fig.~\ref{fig-04}a. Measured by $\Delta
n(r_0)$, the singlet appears about four times more robust than the
polarized state.

These findings can neither be completely explained with the intuition
of non-interacting CFs nor purely on the basis of the numerically
calculated gaps. For the former, we would expect both the $\tf$ and
$\tt$ singlet to have the same response as the Laughlin state. In all
three states all CFs reside in the lowest CF Landau level and they are
excited to the first CF LL by the impurity. On the other hand, the CF
intuition is correct for the $\tt$ and $\tf$ polarized states. 

Regarding the gap energies it is also unexpected that the $\tt$ spin
singlet seems more robust than the polarized state whose gap is
larger. An explanation for the stability of the singlet state must
therefore lie in the structure of the wavefunctions rather than in
the spectrum.

\section{Spin pairing in the $\tt$ singlet state}
\label{sec-4}

The density--density correlation functions provide an alternative
view at the internal structure of a many body state $|\Psi\rangle$. With spin
degree of freedom, there are three distinct types denoted by
$g_{\up\up}$(r), $g_{\dn\dn}(r)$ and $g_{\up\dn}(r)$. 
With $\delta_{\sigma\up}$ being a projector on spin up single-electron
states, the first one is defined as 
$\langle\Psi|\delta(r-r'-r'')\delta_{\sigma'\up}
\delta_{\sigma''\up}|\Psi\rangle$,
the others analogously.

\begin{figure}
\begin{center}
\begin{tabular}{cc}
\includegraphics[scale=.5,angle=0]{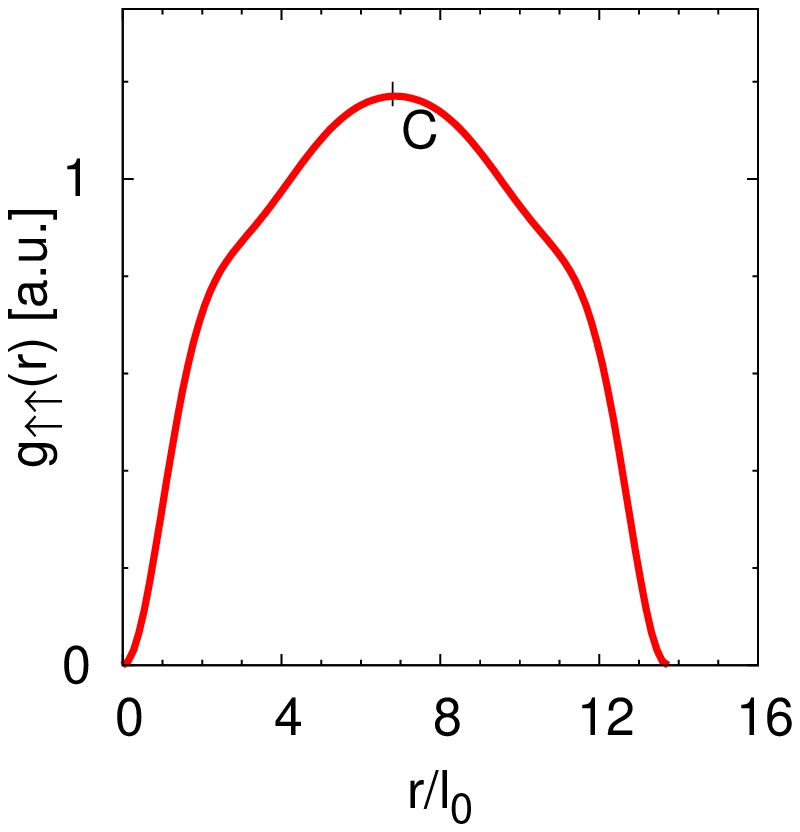}&
\includegraphics[scale=.5,angle=0]{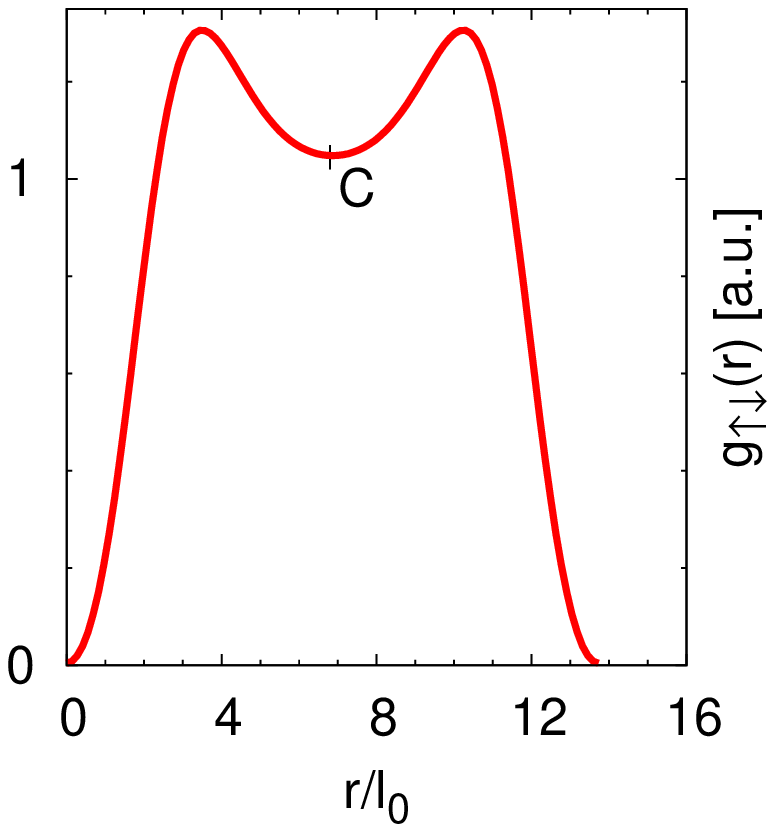}
\end{tabular}
\end{center}
\caption{Correlations between like spins and unlike spins in the
  $\tt$ singlet state. Ten electrons in a square elementary cell (C marks its
  centre), $g(r)$ along its diagonal is shown. Note the maximum in
  $g_{\up\dn}(r)$ around $3.4\ell_0$.}
\label{fig-08}
\end{figure}

The $\nu=\tt$ singlet state displays remarkable structures in these
spin-resolved correlations, Fig.~\ref{fig-08}. While $g_{\up\up}(r)$ has a
pronounced shoulder around $r\approx 2\ell_0$, the unlike spins have a strong
correlation maximum near $3.5\ell_0$. Even though not shown in
Fig.~\ref{fig-08}, this was checked for several different system sizes and the
above dimensions turned out to be always the same. Note that these structures
are significantly stronger than in the case\cite{kamilla:11:1997,vyborny:2005}
of the Laughlin state at $\nu=\ot$.  We interpret this as a signature of spin
pairing: two electrons with opposite spin form an object of the characteristic
size of approximatelly three magnetic lengths which is the mean interparticle
separation $\sqrt{2\pi/\nu}\ell_0\approx 3.1\ell_0$. Qualitatively the same
behaviour was found in the $\tf$ singlet state\cite{vyborny:2005}.

\begin{figure}
\begin{center}
\begin{tabular}{cc}
\includegraphics[scale=.5,angle=0]{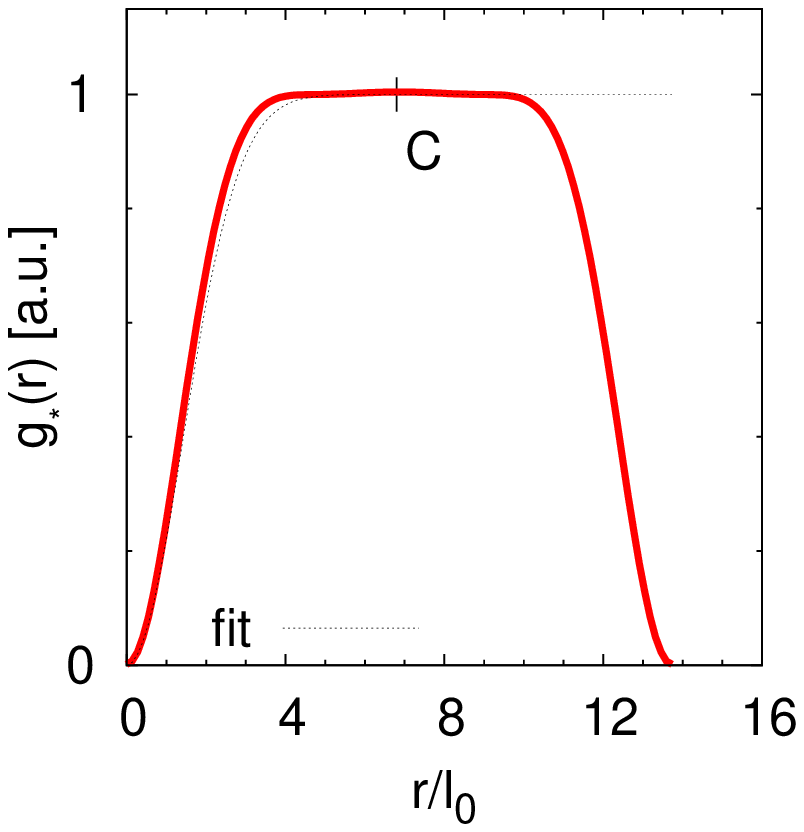}&
\kern-2cm\raise-.5cm\hbox{\includegraphics[scale=.6]{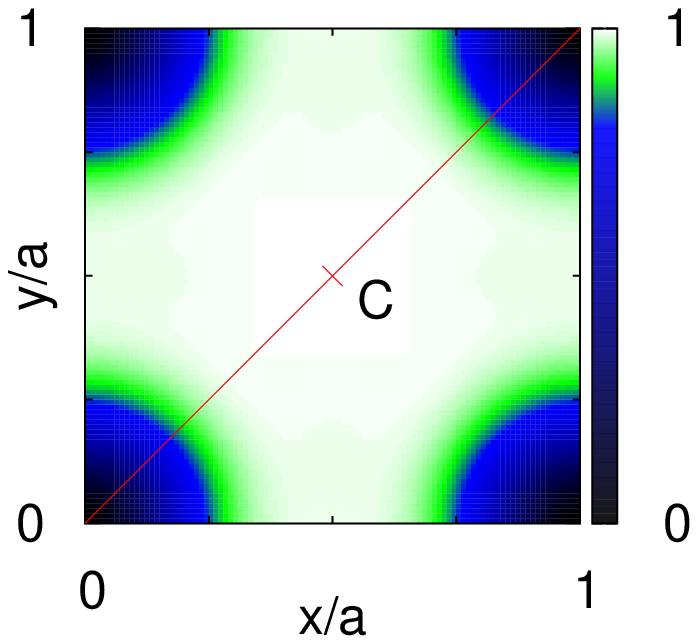}}
\end{tabular}
\end{center}
\caption{Data of Fig.~\ref{fig-08} combined into a spin-unresolved
  correlation function of the $\tt$ singlet state closely resemble the
  completely filled lowest Landau level with $N_e/2$ particles (the fit). The
  $g_*(r)$ in the whole elementary cell and a section along the diagonal are
  shown.}
\label{fig-07}
\end{figure}

Moreover, combining the correlation functions above into a
spin-unresolved one\cite{comm:04}, $g_*(r)=g_{\up\up}+g_{\up\dn}$, we
arrive at a result strongly reminiscent of the completely filled
lowest Landau level (i.e. the ground state at $\nu=1$),
Fig.~\ref{fig-07}. Such a state is characteristic by a simple
correlation hole at $r=0$ passing over monotonously to a
constant\cite{girvin:07:1999}:
$g_{\nu=1}(r)=1-\exp(-r^2/2\ell_0^2)$. Surprisingly, however, the
correlation hole found in $g_*(r)$ corresponds to $g_{\nu=1}$ with
$\ell_0^2$ replaced by $2\ell_0^2$ with a rather good precision (the
fit in Fig.~\ref{fig-07}).

In a finite system with $N_e$ electrons, $g_*=g_{\nu=1}$ would mean
that these electrons form the standard $\nu=1$ state and behave as if
they felt $N_m=N_e$ magnetic flux quanta. With the replacement
$\ell_0^2\to 2\ell_0^2$ we conclude that the electrons feel only
$N_m=N_e/2$ flux quanta, or that each pair of electrons feels one flux
quantum. Based on observations in Fig.~\ref{fig-08},\ref{fig-07} we
suggest that the electrons in the $\tt$ singlet state form 
pairs of total spin zero with characteristic size of $3\ell_0$ and
these pairs condense into a state resembling the completely filled
lowest Landau level. 

The observed robustness of the singlet states against the potential
impurities thus may be related to the robust incompressibility of the filled
lowest Landau level, possibly helped also by the fact that the relevant
particles are not single electrons but rather electron pairs.
We now proceed with the investigation of the spin singlet
states keeping in mind this picture providing a guidance at the
interpretation.

\subsection{Magnetic impurity (MI)}

\begin{figure}
\begin{tabular}{cc}
2/3 & 2/5 \\
\includegraphics[scale=.5]{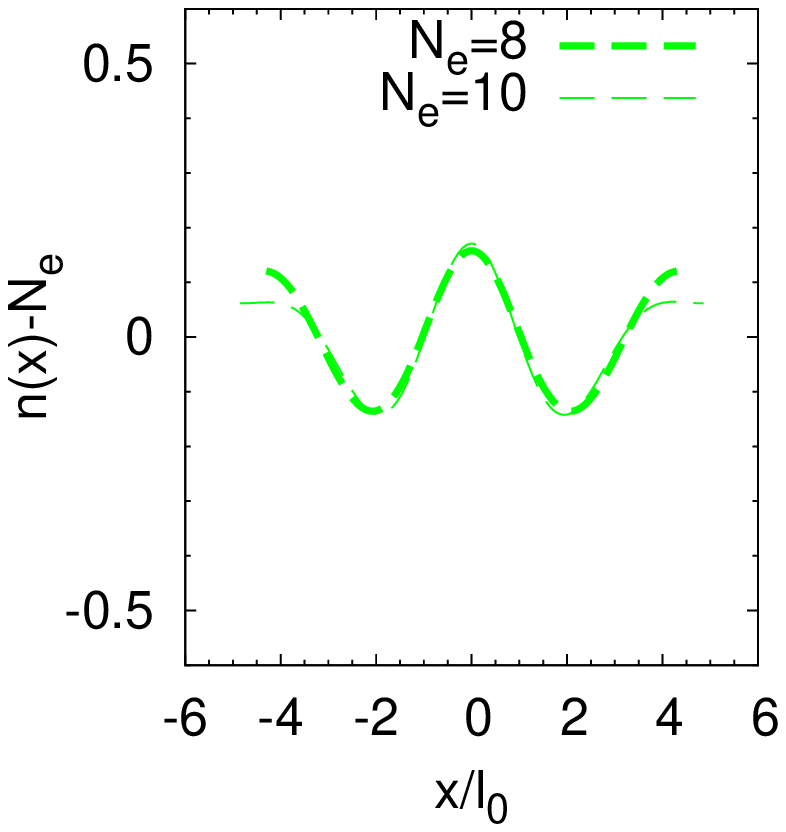} &
\hskip-.5cm\includegraphics[scale=.5]{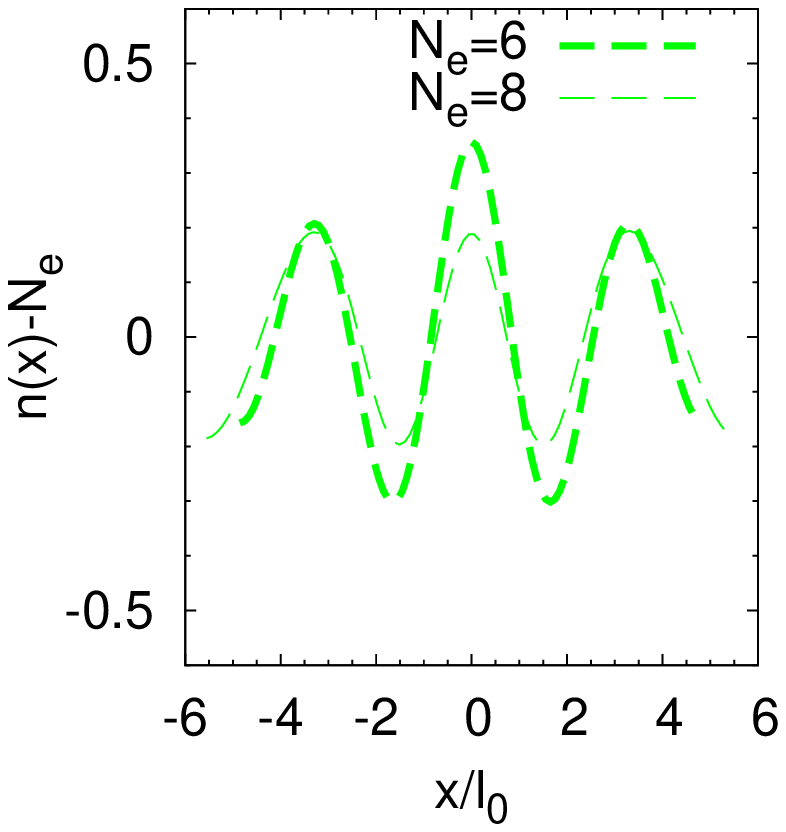} \\
(a) & (b) \\
\includegraphics[scale=.5]{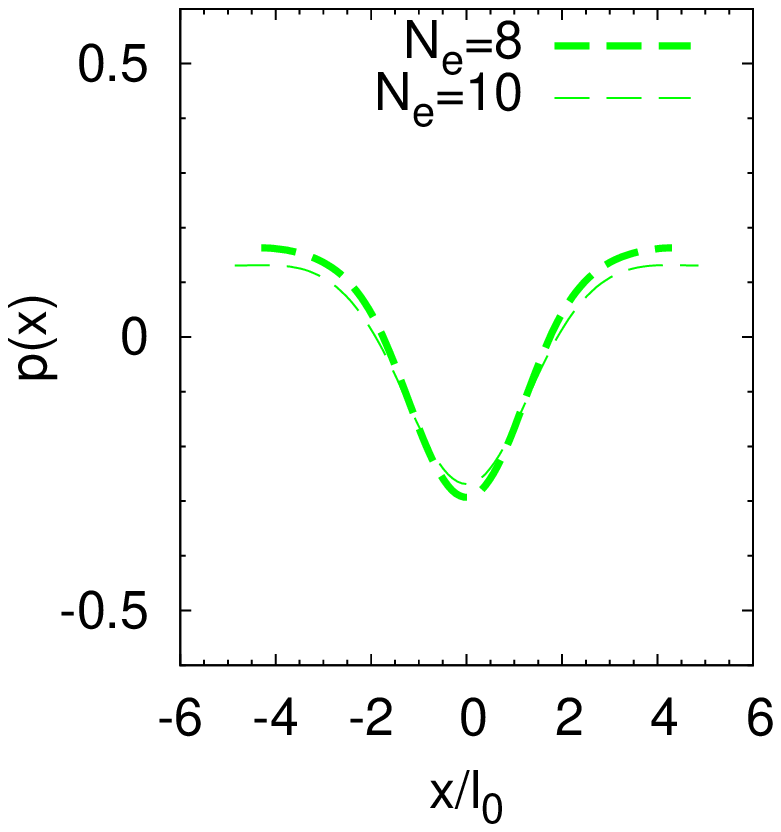} &
\hskip-.5cm\includegraphics[scale=.5]{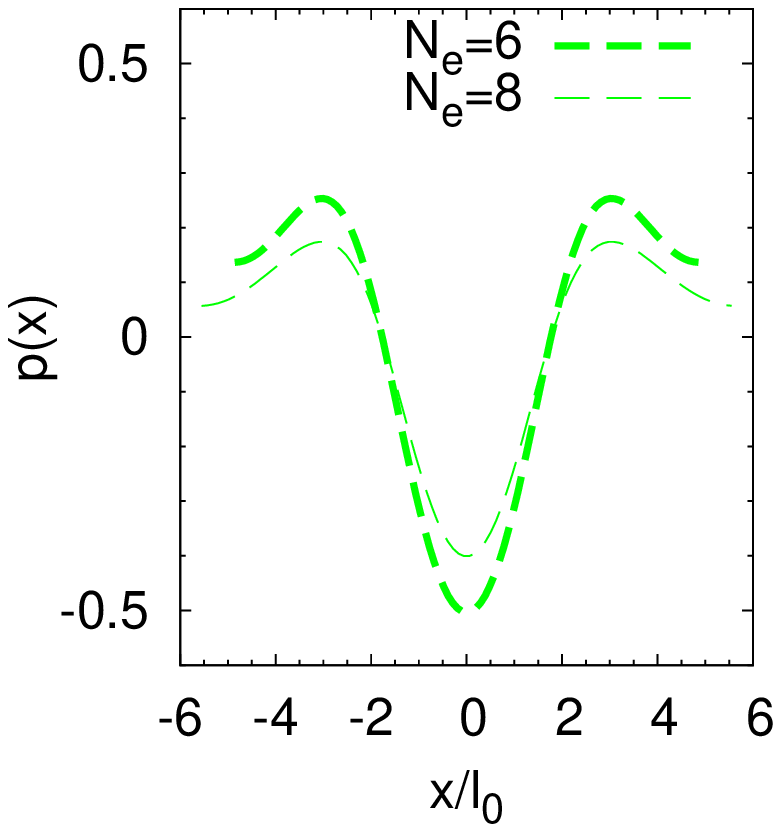} \\
(c) & (d)
\end{tabular}
\caption{Density (a,b) and polarization (c,d) response of the $\tt$ and $\tf$
  singlet states to a magnetic impurity.}
\label{fig-05}
\end{figure}

A magnetic impurity creates a relatively strong spin polarization
$p(x)=[n_\up(x)-n_\dn(x)]/n(x)$ of the
singlet states, Fig.~\ref{fig-05}c,d. The density of
electrons with spin $\up$ at $x=0$ decreases by $40\%$ and $30\%$ for
$\tf$ and $\tt$, respectively. On the other hand, a potential
impurity of the same strength $V_0$ changes the total density of these
states only by $5\%$ and $1.5\%$, Fig.~\ref{fig-04}.

This observation is compatible with the concept of singlet pairs forming a
$\nu=1$ state ($\tt$ singlet). While the potential impurity forces the whole
unwieldy pairs to rearrange, it is relatively easy to polarize the pairs
without moving their centre of mass. In terms of the many-body states, the
weak density response suggests that the correlations in the ground state are
similar to those of low lying excited states.

Several remarks should be added.  (a) Magnetic impurity changes not only the
polarization but also the density, Fig.~\ref{fig-05}a,b. This effect is
proportional to $V_0^2$ and it is a consequence of that the density and spin
density operators do not commute within the lowest Landau
level\cite{girvin:07:1999}. Regardless of the sign of $V_0$, the density
always increases at the impurity position.  (b) The polarization profiles show
only one node in our finite systems, Fig.~\ref{fig-05}c,d. Characteristic
length related to screening of magnetic impurities, if present at all, is
therefore probably rather large.  (c) The $\tf$ singlet state with a magnetic
impurity is the only case in this work where increasing the system size leads
to a notably smaller density (and also polarization) response.

\subsection{Delta plus impurity (DP)}

\begin{figure}
\begin{tabular}{cc}
\hskip-.0cm\includegraphics[scale=.52]{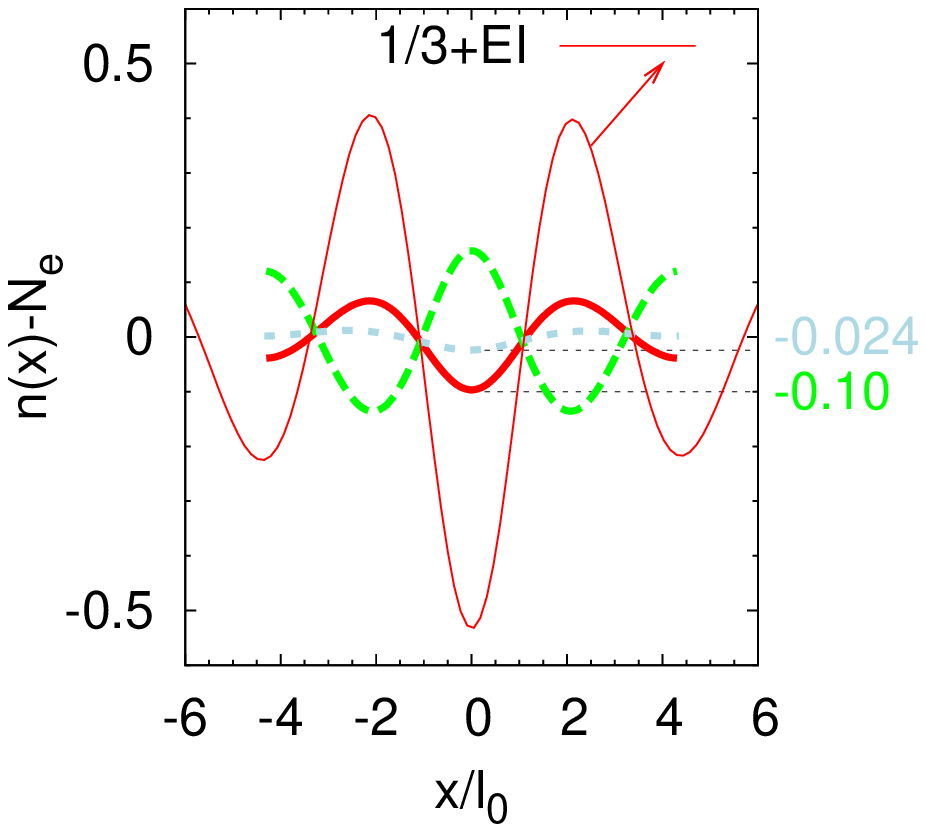}&
\hskip-2.5cm\includegraphics[scale=.52]{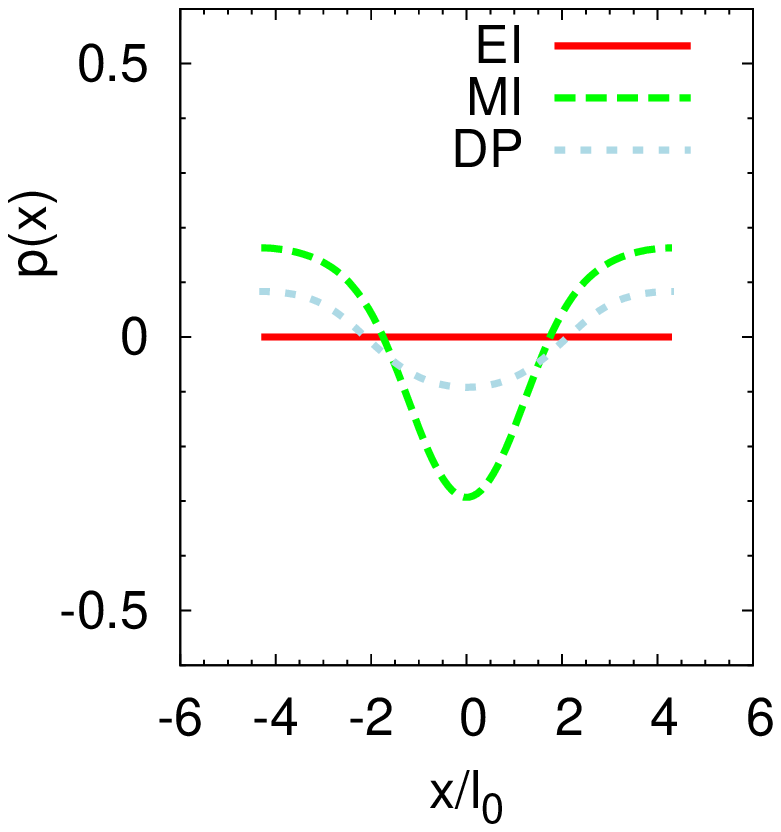} \\
\hskip-1cm (a) & \hskip-2cm(b) 
\end{tabular}
\caption{Density (a) and polarization (b) response to
  all three types (Eq.~\ref{eq-01}) of $\delta$-line impurities acting on the
  $\tt$ singlet ($N_e=8$). Density response of the Laughlin state to a
  potential impurity is shown for comparison in (a) by the thin line.}
\label{fig-06}
\end{figure}

Finally we investigate the $\tt$ singlet state subject to the 'delta plus'
impurity, Eq.~\ref{eq-01}, and found weaker responses than for other types of
impurities, Fig.~\ref{fig-06}. Put into the context of our singlet-pair
incompressible state this finding reflects the strong correlations within the
pair. 

We point out that the response to a DP impurity cannot be derived only from
the knowledge of responses to a potential and to a magnetic impurity even
though EI+MI=2$\times$DP on the level of the Hamiltonian, Eq.~\ref{eq-01}. We
find that the density responses to a potential impurity (EI) and the DP
impurity are in ratio $\approx 4:1$, Fig.~\ref{fig-06}a while the polarization
responses to the magnetic impurity (MI) and the DP are $\approx 3:1$,
Fig.~\ref{fig-06}b. The both being more than $2:1$ means that the capability
of the individual spin species to answer to external perturbations is
suppressed, hence suggesting that their mutual correlation is strong. It is
also worth of recalling that the response of the $\tt$ singlet state to a DP
is thus more than an order of magnitude smaller than the one of the Laughlin
state, Fig.~\ref{fig-06}a. A naive interpretation of $\nu^*=2$, for example,
as of two independent $\nu^*=1$ systems (corresponding to Laughlin
wavefunctions) with different spins therefore fails utterly.

\section{Summary}

We find that the polarized $\nu=\ot$, $\tt$ and $\tf$ states all
respond similarly to isolated impurities, represented in this paper by
a weak $\delta$-line impurity. Since $\tf$ gives slightly
different results than the other two systems we conclude that the
particle-hole conjugation ($\ot$ and $\tt$) is a stronger link than
the reversal of the effective field in the composite fermion picture
($\tt$ and $\tf$).

The singlet states react differently: both $\tt$ and $\tf$ respond
quite unequally and more weakly to an electric potential impurity than
$\ot$. In particular, $\tt$ gives a much weaker response than
$\ot$. This was unexpected because the Laughlin state has the largest
incompressibility gap. The spin-resolved {\em and} spin-unresolved
density-density correlation functions of the $\tt$ singlet state
suggest that electrons in it appear in zero-spin pairs with
characteristic size of $3$ magnetic lengths and these form a
full-Landau-level-like state. High polarizability by magnetic impurities and a
relatively small effect of impurities affecting only one spin species
are compatible with this interpretation.

The authors acknowledge support from the following grants:
AV0Z10100521 of the Academy of Sciences of the Czech Republic (KV),
LC510 of the Ministry of Education of the Czech Republic (KV),
and SFB 508 Quantenmaterialien.


\begin{thebibliography}{10}
\bibitem{comm:04}
More precisely, we used 
$g_*=g_{\up\up}/N_\up(N_\up-1)+g_{\up\dn}/N_\up N_\dn$
to account for the finite size of the system. $N_\up$ and $N_\dn$ is the
number of electrons with spin up and spin down, respectively.
\end{thebibliography}

\end{document}